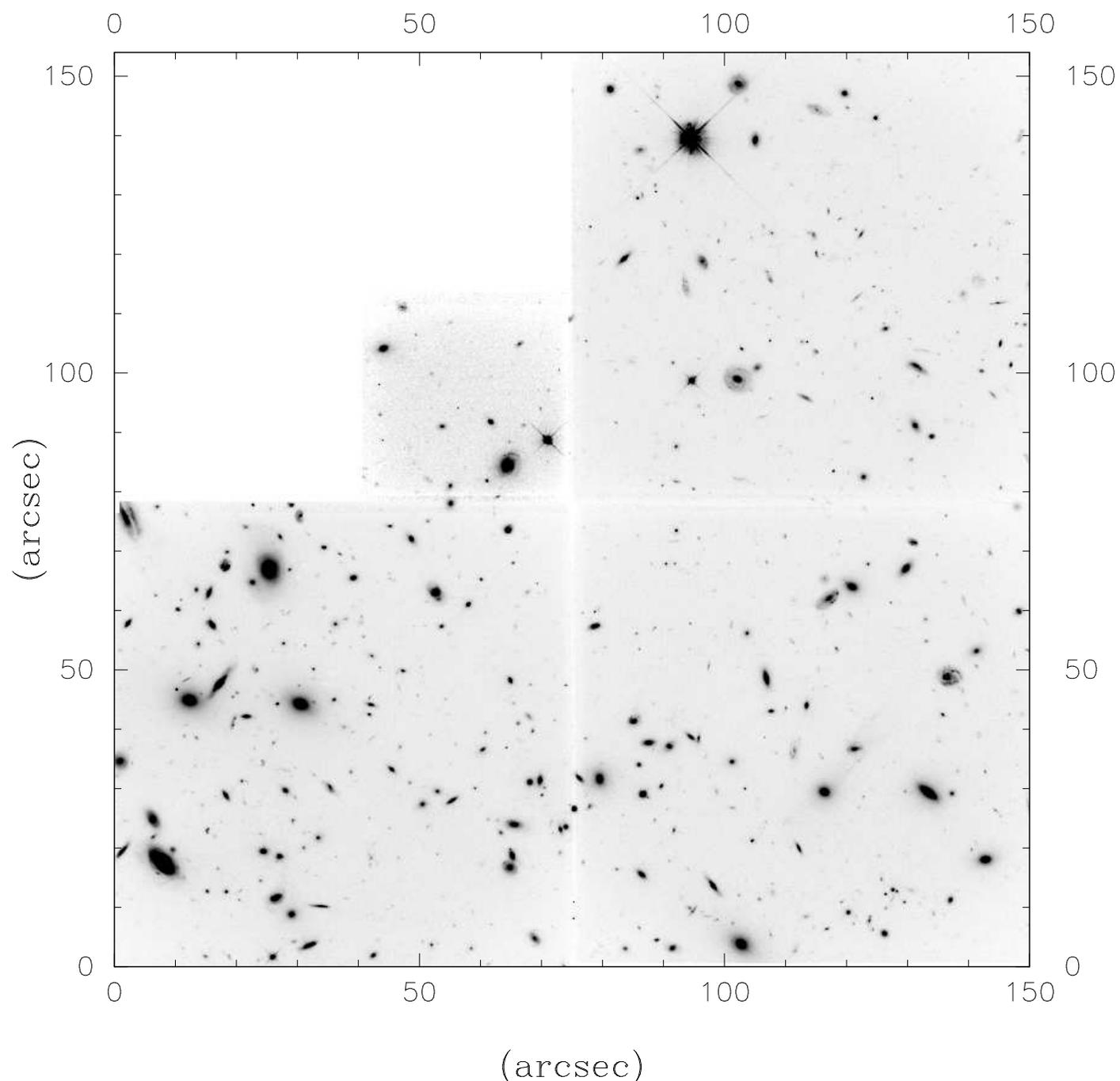

# Cosmological Applications of Gravitational Lensing

Peter Schneider

Max-Planck-Institut für Astrophysik, Postfach 1523, D-85740 Garching, Germany

**Abstract:** The last decade has seen an enormous increase of activity in the field of gravitational lensing, mainly driven by improvements of observational capabilities. I will review the basics of gravitational lens theory, just enough to understand the rest of this contribution, and will then concentrate on several of the main applications in cosmology. Cluster lensing, and weak lensing, will constitute the main part of this review.

## 1 Introduction

Gravitational light deflection has been one of the key tests of Einstein's Theory of General Relativity. Several authors in the 1920's have pointed out that this effect may give rise to spectacular effects, such as multiple images or ring-like images of distant sources, but no one expressed his vision so clearly as Zwicky in 1937, when he claimed that the observation of the gravitational lens effect will be 'a certainty'; he also estimated the probability of a distant source to be multiply imaged to be a few tenth of a percent, very close to modern estimates, and he predicted that the lens effect will allow the determination of the mass of distant cosmic objects and, due to the magnification effect, allow deeper looks into the universe (for an account of the history of this field and for references, see Chap. 1 of Schneider, Ehlers & Falco 1992, hereafter SEF). These predictions were eventually verified when Walsh, Carswell & Weymann (1979) discovered the first lensed QSO, where two QSO images with redshift $z_s = 1.41$, separated by $6''$, have nearly identical spectra from radio to X-ray frequencies, with a giant elliptical galaxy at redshift $z_d = 0.36$, situated in a cluster of galaxies, between the images. Today, the number of multiply-imaged QSOs is about 15; in addition, 6 ring-shaped radio images have been found, in some cases with a (lower-redshift) galaxy at the ring center (for a recent review of the observational situation, see Refsdal & Surdej 1994). The discovery of giant luminous arcs in 1986 by Lynds & Petrosian (1986) and Soucail et al. (1987) has shown that clusters of galaxies can act as efficient lenses; cluster lensing today is one of the most active fields of gravitational lensing (for a recent review, see Fort & Mellier 1994). Finally, the impressive demonstration (Alcock et al. 1993, Aubourg et al. 1993, Udalski et al. 1993) of the feasibility of the suggestion by Paczyński (1986) to search for compact dark objects in the halo of our Galaxy, has led to an active and successful search of Galactic microlensing events, both towards the LMC and the Galactic bulge (for a recent review, see Paczyński 1996).

These discoveries have opened up a new road towards investigating massive structures in the universe. Since gravitational light deflection is insensitive to the nature and physical state of the deflecting mass, it is ideally suited to study dark matter in the universe. In this review, only some aspects of this exciting research field can be treated; whereas strong lensing applications will be discussed in Sect. 3, I will describe cluster lensing and weak lensing in Sect. 4 in somewhat more detail. However, the necessary tools must be prepared, which will be done in Sect. 2.

## 2 Lensing geometry

### 2.1 The lens equation

The formal description of gravitational lensing is basically simple geometry. Consider a mass distribution (the deflector) at some distance $D_\mathrm{d}$ from us, and some source at distance $D_\mathrm{s}$ (see Fig. 1). Then, draw a reference line ('optical axis') through lens and observer, define planes ('lens plane' and 'source plane') perpendicular to this optical axis through lens and source, and measure the transverse separations of a light ray in the source and lens plane by $\boldsymbol{\eta}$ and $\boldsymbol{\xi}$, respectively. Then from simple geometry, the relation between these two vectors is

$$\boldsymbol{\eta} = \frac{D_\mathrm{s}}{D_\mathrm{d}} \boldsymbol{\xi} - D_\mathrm{ds} \hat{\boldsymbol{\alpha}}(\boldsymbol{\xi}) \quad , \tag{1}$$

where $\hat{\boldsymbol{\alpha}}(\boldsymbol{\xi})$ is the *deflection angle*. Since all deflection angles one is interested in are very small (even in clusters of galaxies, the deflection angles are well below $1'$), and thus the gravitational fields are weak, the *linearized field equation of General Relativity* can be employed, which implies that the deflection angle is a linear functional of the mass distribution. Since the deflection angle of a light ray passing a point mass $M$ at separation $r$ is $4GM/(rc^2)$, the deflection angle at position $\boldsymbol{\xi}$ caused by a mass distribution descibed by the *surface mass density* $\Sigma(\boldsymbol{\xi})$ becomes

$$\hat{\boldsymbol{\alpha}}(\boldsymbol{\xi}) = \int_{\mathbb{R}^2} d^2\xi' \, \frac{4G\Sigma(\boldsymbol{\xi}')}{c^2} \, \frac{\boldsymbol{\xi} - \boldsymbol{\xi}'}{|\boldsymbol{\xi} - \boldsymbol{\xi}'|^2} \quad , \tag{2}$$

where the integral extends over the lens plane.

The simple description of a gravitational lens situation can be justified much more thoroughly from Relativity; the reader is referred to SEF, Chap. 4, and Seitz, Schneider & Ehlers (1994) for a rigorous treatment. Here it suffices to note that for all situations encountered in this review, the gravitational lens equations provide excellent approximations; in particular, the simple geometrical derivation of (1) remains valid in a Friedmann–Lemaître universe if the distances are interpreted as *angular-diameter distances*.

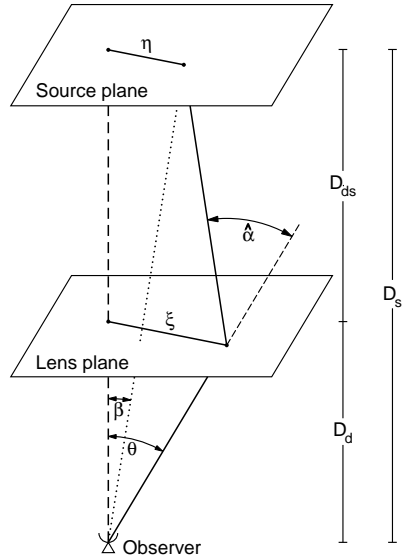

**Fig. 1.** The geometry of a gravitational lens

It is convenient to replace the physical lengths in (1) by angular variables, by defining $\boldsymbol{\beta} = \boldsymbol{\eta}/D_{\mathrm{s}}$, $\boldsymbol{\theta} = \boldsymbol{\xi}/D_{\mathrm{d}}$,

$$\boldsymbol{\alpha}(\boldsymbol{\theta}) = \frac{D_{\mathrm{ds}}}{D_{\mathrm{s}}}\hat{\boldsymbol{\alpha}}(D_{\mathrm{d}}\boldsymbol{\theta}) = \frac{1}{\pi}\int_{\mathbb{R}^2} \mathrm{d}^2\theta'\, \kappa(\boldsymbol{\theta}')\frac{\boldsymbol{\theta}-\boldsymbol{\theta}'}{|\boldsymbol{\theta}-\boldsymbol{\theta}'|^2} \quad, \tag{3}$$

with the *dimensionless surface mass density*

$$\kappa(\boldsymbol{\theta}) = \frac{\Sigma(D_{\mathrm{d}}\boldsymbol{\theta})}{\Sigma_{\mathrm{cr}}} \quad \text{with} \quad \Sigma_{\mathrm{cr}} = \frac{c^2}{4\pi G}\frac{D_{\mathrm{s}}}{D_{\mathrm{ds}}D_{\mathrm{d}}} \quad; \tag{4}$$

then the *lens equation* simply reads

$$\boldsymbol{\beta} = \boldsymbol{\theta} - \boldsymbol{\alpha}(\boldsymbol{\theta}) \quad. \tag{5}$$

The *critical surface mass density* $\Sigma_{\mathrm{cr}}$ is a characteristic value which separates strong from weak lenses; if $\kappa \ll 1$ everywhere (i.e., $\Sigma \ll \Sigma_{\mathrm{cr}}$), then the deflector is weak, whereas if $\kappa \sim 1$ for some $\boldsymbol{\theta}$, the lens may produce multiple images and is called strong. *Multiple images* occur if the lens equation (5) has multiple solutions $\boldsymbol{\theta}$ for the same source position $\boldsymbol{\beta}$.

### 2.2 The deflection potential, and the time-delay

Using the identity $\nabla \ln |\mathbf{x}| = \mathbf{x}/|\mathbf{x}|^2$, one sees that the deflection angle $\boldsymbol{\alpha}$ can be written as the gradient,

$$\boldsymbol{\alpha}(\boldsymbol{\theta}) = \nabla \psi(\boldsymbol{\theta}) \quad \text{with} \quad \psi(\boldsymbol{\theta}) = \frac{1}{\pi} \int_{\mathrm{I\!R}^2} \mathrm{d}^2 \theta' \, \kappa(\boldsymbol{\theta}') \ln |\boldsymbol{\theta} - \boldsymbol{\theta}'| \quad , \qquad (6)$$

of the *deflection potential* $\psi$. If we define the *Fermat potential*

$$\phi(\boldsymbol{\theta}; \boldsymbol{\beta}) := \frac{|\boldsymbol{\theta} - \boldsymbol{\beta}|^2}{2} - \psi(\boldsymbol{\theta}) \;, \quad \text{then} \quad \nabla \phi(\boldsymbol{\theta}; \boldsymbol{\beta}) = 0 \qquad (7)$$

is equivalent to the lens equation (5). In fact, one can show that $\phi(\boldsymbol{\theta}; \boldsymbol{\beta})$ is, up to an affine transformation, the *light travel time* along a light ray from the source at $\boldsymbol{\beta}$ via a point $\boldsymbol{\theta}$ in the lens plane to the observer. Hence, (7) expresses the fact that physical light rays are those for which the light travel time is stationary – which is *Fermat's principle* in gravitational lens theory.

If a source has multiple images, the light travel time along the different rays will be different. From the interpretation of $\phi$ it is clear that the *time delay* $\Delta t$ is proportional to the difference of the Fermat potential at the image positions. One finds:

$$c \, \Delta t(\boldsymbol{\beta}) = \frac{D_\mathrm{d} D_\mathrm{s}}{D_\mathrm{ds}} (1 + z_\mathrm{d}) \left[ \phi\left(\boldsymbol{\theta}^{(1)}, \boldsymbol{\beta}\right) - \phi\left(\boldsymbol{\theta}^{(2)}, \boldsymbol{\beta}\right) \right] \quad . \qquad (8)$$

### 2.3 Magnification and image distortion

Light bundles are not only deflected as a whole, but differential deflection occurs. Hence, in a first approximation, a circular light bundle aquires an elliptical cross section after passing a deflector. The differential deflection changes the solid angle subtended by a source. Since the *surface brightness* (or the specific intensity) is unchanged by light deflection – this follows from Liouville's theorem, or the fact that light deflection neither creates nor destroys photons – the change in solid angle leads to a change of observed flux from a source: the flux of an infinitesimally small source with surface brightness $I$ and solid angle $\Delta\omega$ is $S = I \, \Delta\omega$. If $\Delta\omega_0$ is the solid angle subtended by an infinitesimally small source in the absence of a deflector, then the observed flux of an image of this source at $\boldsymbol{\theta}$ is $S = \mu(\boldsymbol{\theta}) \, S_0$, where the *magnification* $\mu$ of an image of an infinitesimally small source is

$$\mu(\boldsymbol{\theta}) = |\det A(\boldsymbol{\theta})|^{-1} \;, \quad \text{where} \quad A(\boldsymbol{\theta}) = \frac{\partial \boldsymbol{\beta}}{\partial \boldsymbol{\theta}} \qquad (9)$$

is the Jacobian matrix of the lens equation[1]; in components, $A_{ij} = \partial \beta_i / \partial \theta_j \equiv \beta_{i,j}$. The matrix $A$ describes the *locally linearized lens mapping*. Note that $\operatorname{tr} A(\boldsymbol{\theta}) = 2[1 - \kappa(\boldsymbol{\theta})] = 2 - \nabla^2 \psi(\boldsymbol{\theta})$, i.e., the deflection potential $\psi$ satisfies a *Poisson-like equation.* The fact that the two eigenvalues of $A$ will be different in general implies that a circular source will be imaged, to first approximation, into an ellipse. We can write the components of $A$ as

$$A = \begin{pmatrix} 1 - \kappa - \gamma_1 & -\gamma_2 \\ -\gamma_2 & 1 - \kappa + \gamma_2 \end{pmatrix} = (1 - \kappa)\mathcal{I} - |\gamma| \begin{pmatrix} \cos(2\varphi) & \sin(2\varphi) \\ \sin(2\varphi) & -\cos(2\varphi) \end{pmatrix} \ , \tag{10}$$

where $\gamma$ is called *shear* and describes the tidal gravitational forces ($\mathcal{I}$ is the two-dimensional identity matrix). The components of the shear are given by second partial derivatives of the deflection potential,

$$\gamma_1 = \frac{1}{2}(\psi_{,11} - \psi_{,22}) \ , \ \gamma_2 = \psi_{,12} \ , \ \kappa = \frac{1}{2}(\psi_{,11} + \psi_{,22}) \ . \tag{11}$$

---

[1] The *magnification of a source* is then the sum of the magnifications of its images; the *magnification* of an extended source is the surface-brightness averaged magnification of its source points.

The eigenvalues of $A$ are $1 - \kappa \pm |\gamma|$, where $|\gamma| = \sqrt{\gamma_1^2 + \gamma_2^2}$, the axis ratio of the elliptical image of a circular source is given by the ratio of these two eigenvalues, and the orientation of the major axis is described by the angle $\varphi$. We shall later discuss the image distortion for a general source.

Note that $\det A$ can vanish, which formally implies a diverging magnification. Of course, real magnifications remain finite. A real source is extended, and the magnification averaged over an extended source is always finite. Even if we had a point source, the magnification would remain finite: in this case, the geometrical optics approximation breaks down and light propagation had to be described by wave optics, yielding finite magnifications (see Chap. 6 of SEF). Astrophysically relevant situations involve sufficiently large sources for the geometrical optics approximation to be valid. The closed curves on which $\det A = 0$ are called *critical curves*; the corresponding curves in the source plane, obtained by inserting the critical points into the lens equation, are called *caustics*. An image close to a critical curve can have a large magnification; also, the number of images of a source changes by $\pm 2$ if and only if the source position changes across a caustic. In this case, two images merge at the corresponding point of the critical curve, thereby brightening, and disappear once the source has crossed the caustic. The caustic is not necessarily a smooth curve, but it can develop *cusps*. A source close to, and inside a cusp has three bright images close to the corresponding point of the critical curve, whereas it has one bright image if situated just ouside the cusp.

## 3 Applications: Strong lensing

In this section I will discuss some of the cosmological applications of gravitational lensing which are related to galaxy-sized deflectors and those of smaller mass, keeping cluster-size lenses for the next section. The list presented here is of course non-exhaustive; I refer the interested reader to the review by Blandford & Narayan (1992) and the other reviews mentioned in the introduction.

### 3.1 Mass determination

The perhaps most obvious application of gravitational lensing is the determination of the mass of the deflector. The simplest situation in which a mass can be determined is that of a spherical deflector, with a source right behind the lens' center. If the lens is sufficiently strong, the source will form a ring-shaped image ('Einstein ring'), of which several examples have been found. For an axi-symmetric mass distribution, the deflection angle becomes $\hat{\alpha}(\theta) = 4GM(<\theta)/(c^2 D_{\mathrm{d}}\theta)$, and so the lens equation, with the source at the origin, reads $\theta D_{\mathrm{s}} = \hat{\alpha}(\theta) D_{\mathrm{ds}}$. Combining the last two equations, one finds

$$M(<\theta) = \pi \left(D_{\mathrm{d}}\theta\right)^2 \Sigma_{\mathrm{cr}} \quad . \tag{12}$$

Hence, the mean surface mass density inside the Einstein ring is the critical surface mass density, and thus the mass inside the Einstein ring can be determined once its angular diameter and the redshifts of lens and source are measured.

In fact, even if no ring-shaped image is observed, a mass estimate based on the preceding ideas is often useful and surprisingly accurate. For example, a quadruple image system allows to trace approximately the Einstein 'circle', and a mass estimate can be obtained from (12). However, more detailed modelling is warranted in such cases. It should be mentioned that the mass inside the inner $0\rlap{.}''9$ of the lensing galaxy in the quadruple QSO 2237+0305 (the so-called 'Einstein cross') has been determined with an accuracy of a few percent (Rix, Schneider & Bahcall 1992), with the largest uncertainty being due to the Hubble constant. For modelling extended images, such as radio rings, elaborate techniques have been developed (Kochanek & Narayan 1992) and successfully been applied (Kochanek et al. 1989; Kochanek 1995a; Chen, Kochanek & Hewitt 1995; Wallington, Kochanek & Narayan 1995).

Whereas the mass determination from strong lensing events is the most accurate extragalactic mass determination (again: this method does not depend on the nature or state of the matter), the limitations of this method should be kept in mind: it measures the mass inside 'cylinders', i.e., the projected mass, and it measures the mass only in the inner part of a lensing galaxy.

### 3.2 Measuring the Hubble constant

Refsdal (1964) pointed out that a gravitational lens system can be used to determine the Hubble constant. The basic argument is as follows: all observables in a gravitational lens system are dimensionless (angles, flux ratios – although fluxes are measured, they provide no constraint on the geometry since the intrinsic luminosity of the source is unknown –, redshifts etc.), except the time delay between any pair of images. Now consider the size of the universe to be scaled by a factor $L$; then, all dimensionless observables were unchanged, but the time delay would also change by a factor $L$. Thus, a measurement of the time delay enables one to determine the absolute size of the lensing geometry, and thus the Hubble constant.

From (8) we see that the time delay can be factorized as follows:

$$\Delta t = \frac{1}{H_0} F \text{ (lens model}, z_\mathrm{d}, z_\mathrm{s}, \text{cosmology)} \quad . \tag{13}$$

The dimensionless function $F$ depends on the cosmological parameters $\Omega$ and $\Lambda$, but this dependence is not very strong if the source and lens redshifts are smaller than $\sim 2$ and $\sim 0.5$, respectively. The redshifts of source and lens are assumed to be known. The largest uncertainty is the construction of a reliable lens model; we shall discuss this further below.

The second problem which occurs is the measurement of the time delay itself. For the double QSO 0957+561, monitoring of the two QSO images has been done in the optical (e.g., Vanderriest et al. 1989, Schild & Thomson 1995) and the radio (Roberts et al. 1991, Haarsma et al. 1996) wavebands for over 15 years. Despite this enormous observational effort, there has been no agreement on the value of $\Delta t$, with values between 410 days and 540 days occurring in the literature, because: (i) the QSO has not been very cooperative, i.e., it has not varied strongly in the last 15 years; (ii) some variability of the images must be attributed to *microlensing*[2]; (iii) the QSO is observable from the ground with optical telescopes for only 8 months a year, so that the lightcurves have gaps; this does not apply to the radio lightcurves, but due to the changing configurations of the VLA, the radio lightcurves also have gaps. A cross-correlation of the two lightcurves is thus subject to windowing effects. Furthermore, data points with underestimated errors can affect the resulting time delay and thus require the usage of robust statistical methods (for a thorough discussion of these issues, see Press, Rybicki & Hewitt 1992, Pelt et al. 1994).

Nevertheless, even if the time delay in 0957+561 is measured, its use for the determination of the Hubble constant will be limited, due to the uncertainty of the lens model. The large angular separation of this system ($\sim 6''$) implies that the image splitting is caused by a combination of the main (elliptical) galaxy at $z_\mathrm{d} = 0.36$ and a cluster in which that galaxy is embedded; in addition, there is a second concentration of galaxies in the field, at a redshift of $z \sim 0.5$. The description of the mass distribution thus requires more parameters than available constraints from the observations, leaving a large freedom for the function $F$ in (13) (see, e.g., Bernstein, Tyson & Kochanek 1993). In addition, if $\kappa_0(\boldsymbol{\theta})$ describes a mass distribution for the lens which is compatible with all observational constraints [image positions, relative magnification matrix $A(\boldsymbol{\theta}^{(1)}) A^{-1}(\boldsymbol{\theta}^{(2)})$],

---

[2] Since the matter in the lensing galaxy consists partly of stars, the mass distribution is grainy; the emitting region of the optical continuum light of QSOs is sufficiently small to be sensitive to the gravitational field of stars in the lensing galaxy, down to about Jupiter mass. Whereas the stellar gravitational field does not noticibly affect the angular position of the QSO images, it affects the magnifications, and thus the flux of the images. This effect has been clearly observed in the quadruple QSO 2237+0305 (Houde & Racine 1994, and references therein), as in this system the fluxes of the four QSO images vary independently, whereas any intrinsic variation of the QSO must show up in all four images within the expected time delay of $\sim 1$ day. Note that this microlensing has led to interesting upper bounds on the size of the QSO emitting region (Rauch & Blandford 1991; Jaroszyński, Wambsganss & Paczyński 1992)

then the whole family $\kappa(\boldsymbol{\theta}) = \lambda\kappa_0(\boldsymbol{\theta}) + (1-\lambda)$ of mass distributions satisfies the observational constraints, but the function $F$ in (13) scales like $\lambda$, thus affecting the resulting value of $H_0$ (Gorenstein, Falco & Shapiro 1988). This *mass sheet degeneracy* is always present, but is particularly severe in a case like 0957+561 where the presence of a mass sheet is in fact concluded from the presence of a cluster. The mass sheet degeneracy then implies that gravitational lensing can strictly yield only upper bounds on the Hubble constant.

Perhaps the most promising system currently known for the determination of $H_0$ is the Einstein ring B0218+35.7 (Patnaik et al. 1993), which contains two compact flat-spectrum image components. These compact components are expected to vary, thus enabling the measurement of the time delay, whereas the ring can be used to construct a detailed lens model. Since an extended image yields much more information about the lensing geometry than multiply imaged point-like sources, this system will be much better for constraining the function $F$ in (13), also because the small image separation ($0\rlap{.}''35$) points towards lensing by an isolated (spiral) galaxy. Furthermore, the compact radio components are sufficiently extended (they have been resolved with VLBA observations – see Patnaik, Porcas & Browne 1995) as to not be affected by microlensing. Indeed, from the variability of the *polarized* flux, a preliminary value for the time delay ($\Delta t = 12 \pm 3$ days) has been obtained (Corbett, Browne & Wilkinson 1996).

A value of $H_0$ measured from lensing would be valuable for several reasons: it is a measurement which is completely independent of any local 'distance ladder', it would measure $H_0$ on a truly cosmic scale, and thus being independent of local peculiar velocity fields, and also because an agreement between measurements on cosmic scales with those measured locally would provide a strong support for the validity of standard Friedmann–Lemaître cosmological models.

### 3.3 Galactic microlensing

Among the currently most active fields of lensing research is Galactic microlensing, i.e., lensing by stars in our Galaxy. Paczyński (1986) suggested that a search for such microlensing events may lead to the discovery of, or to an upper limit on the density of compact objects in the halo of our Galaxy, which are dark matter candidates. In this case, stars in the LMC are sources which are lensed by halo objects. As a 'control experiment', he suggested (Paczyński 1991) to observe stars in the Galactic bulge; in this case, the lenses are known to exist, namely the disk stars. The signature of microlensing is a characteristic lightcurve of the lensed star which is described by only four parameters. However, the difficulty of both experiments is the incredibly small lensing probability: about 1 out of $10^{-7}$ stars in the LMC is lensed at any given time if the halo of our Galaxy is made of compact objects. This implies that millions of stars have to be monitored, and the microlensing events have to be extracted from these many lightcurves which include many variable stars. It therefore came as a surprise when three groups announced their detection of microlensing events in the second half of 1993. Today (Oct. 1995), more than hundred microlensing events are known, most of them towards the Galactic bulge (see Paczyński 1996 for a review). The main result of these experiments is that the microlensing rate towards the LMC is smaller than expected, by about a factor of 5, but that the event rate towards the Galactic bulge is larger by a factor of three than expected from naive Galactic mass models. The latter fact is interpreted as indicating that our Galaxy has a bar which is pointing nearly towards us, and that this bar constitutes a major fraction of the microlensing optical depth (Zhao, Spergel & Rich 1995). A variation of the optical depth to microlensing with angular position will allow detailed mass models for the Galaxy. The small microlensing event rate towards the LMC indicates that the halo of the Galaxy is not mainly composed of compact objects, its best-fitting mass fraction being about 20% (Alcock et al. 1995). However, at least part of the lensing optical depth can be provided by objects in the LMC itself or nonhalo Galactic objects.

The incredibly large frequency of publications on galactic microlensing events indicates that this research will continue to yield important results; e.g., on the Galactic mass distribution, the frequency of binary stars, on the dynamics within the Galactic bar, and can even be used to search for planetary systems. It should also be borne in mind that the results from such experiments provide an eldorado for people working on stellar variability!

### 3.4 Lensing statistics and compact dark matter in the universe

The fraction of all high-redshift QSOs which are multiply imaged is proportional to the number density of lenses in the universe; hence, from the observed fraction of multiply imaged QSOs it is possible to constrain the statistical properties of the lens population.

The probability that a QSO is multiply imaged depends on its redshift (the larger the redshift, the more likely is a lens in the line-of-sight), its luminosity (because of the magnification bias[3]), the number density of galaxies (and its possible cosmological evolution), the mass and mass profiles of galaxies, and the cosmological model. Furthermore, the angular separation statistics of the multiple images depends on the masses and redshifts of the lenses, as well as on the cosmological model. The observed angular separation statistics depends furthermore on the observational selection function, which takes into account the finite angular resolution of the observations and the dynamic range of flux ratios which can be observed, depending on the angular separation of the images.

---

[3] QSO samples are flux limited. If a source is magnified, it can enter the flux-limited survey, although its unlensed flux may be below the flux threshold of the sample. Since multiply imaged QSOs are always magnified – typically by a factor of $\sim 4$ for double QSOs, and by a factor $\sim 10$ or higher for quadruple QSOs – multiply imaged QSOs are overrepresented in flux-limited samples. This effect is called magnification bias, and it is larger for steep source counts: the steeper the counts, the more faint QSOs are there for any bright QSO, and thus the reservoir out of which sources can be magnified above the flux threshold is larger. QSO counts are very steep for bright QSOs with $m \lesssim 19$, and flatten considerably for fainter magnitudes (see Hartwick & Schade 1990); hence, the magnification bias is large for bright QSOs.

Several lens surveys have been completed in recent years (for references, see Kochanek 1995b), both in the optical and radio. In order to make use of the magnification bias, and thus to increase the probability that a QSO is multiply imaged, these surveys were performed for the apparently most luminous QSOs, i.e., for bright high-redshift QSOs. For these surveys, the selection function can be reasonably well determined (Kochanek 1993a).

A statistical analysis of the results of these lens surveys consists in a parametrized description of the lens population. Kochanek (1993b) modelled the lensing galaxies as singular isothermal spheres, used a Faber-Jackson relation for the dependence of velocity dispersion (the parameter characterizing the lensing properties of an isothermal sphere) on luminosity, $\sigma/\sigma_* \propto (L/L_*)^\eta$, where $L_*$ is the characteristic luminosity which enters the (Schechter) luminosity function of galaxies. He then used a maximum-likelihood analysis to obtain the best-fitting parameter values from the lens surveys, assuming a constant comoving lens population. A similar analysis was carried out by Maoz & Rix (1993), who investigated also different mass profiles for the lensing galaxies.

The main results of these studies can be summarized as follows: the observed statistics of multiply imaged QSOs is fully compatible with the 'standard assumptions' about the galaxy population and cosmology. The best fit value of $\sigma_*$ is $245 \pm 30$ km/s, very much in agreement with dynamically consistent models of early-type galaxies (spirals, though more numerous, contribute only little to the lensing probability), and the best-fit values for the Faber-Jackson index and the faint-end slope $\alpha$ of the Schechter function are $\eta \sim 4$ and $\alpha \sim -1.1$, again fully compatible with the canonical values. For flat universes with $\Omega + \Lambda = 1$, the best-fit value is $\Lambda = 0$, and a formal upper limit of $\Lambda \leq 0.66$ (95% confidence) can be obtained (Kochanek 1995b). Models in which elliptical galaxies have no dark halo do not reproduce the observed statistics; they predict too few large separation systems.

There is not much room for compact 'dark' lenses with mass in excess of $10^{11} M_\odot$, given that in the majority of the multiple QSOs a (luminous) lens 'between' the images is detected. However, the constraints are less strong for lower-mass objects. For lens masses larger than about $10^6 M_\odot$, these can in principle be detected (or ruled out) with radio-interferometric observations. Kassiola, Kovner & Blandford (1991) analyzed available VLBI observations to put an upper limit of $\Omega_c \leq 0.4$ on the cosmological density of compact objects in the mass range $10^7 M_\odot \lesssim M \lesssim 10^9 M_\odot$; this limit and the corresponding mass range will very soon be dramatically improved, following dedicated VLBI surveys (Augusto, Wilkinson & Browne 1996; Patnaik et al. 1996). The image splitting by lenses with $M \lesssim 10^5 M_\odot$ cannot be resolved even with VLBI; nevertheless, a significant cosmological density of lenses with $M \gtrsim 10^3 M_\odot$ can be ruled out if gamma-ray bursts are at cosmological distances; in that case, lensing of bursts would lead to multiple bursts with delay of $\sim 2 \times 10^{-5} (M/M_\odot)$ seconds (Blaes & Webster 1992), and no obvious candidates for such multiple bursts have been identified yet. Following an early idea of Canizares (1982), Schneider (1993) has obtained upper limits on the density of compact objects in the mass range $3 \times 10^{-4} M_\odot \lesssim M \lesssim 10^{-1} M_\odot$, down to a limit of $\Omega_c \lesssim 0.1$, from constraints on the variability of high-redshift QSOs: a cosmological population of such lenses would lead to the magnification of high-redshift QSOs, and since sources and lenses are moving, the magnification will change in time, leading to lens-induced variability. If the preceding limits on $\Omega_c$ are violated, QSOs would be more variable than observed [the lower mass limit is due to the finite size of QSOs; lenses with $M \lesssim 10^{-4} M_\odot$ cannot magnify the continuum flux of QSOs significantly; the upper mass limit is due to the finite time of observations from which these constraints were obtained – the Hawkins & Véron (1993) sample of variability-selected QSOs]. Since the continuum source of QSOs is much smaller than the broad line region, lenses with $10^{-3} M_\odot \lesssim M \lesssim 10^2 M_\odot$ can magnify the continuum flux, but not the line flux; a cosmologically significant density of compact objects in this mass range would thus lead to small line-to-continuum fluxes of some high-redshift QSOs. The observed lack of this effect has led Dalcanton et al. (1994) to obtain an upper limit of $\Omega_c \lesssim 0.1$ for lenses in the above mentioned mass range.

## 4 Cluster lensing and weak lensing

When giant luminous arcs were first explicitly mentioned by Lynds & Petrosian (1986) and Soucail et al. (1987)[4], they came as a surprise. Whereas alternative explanations for them have been put forward, the redshift determination of the arc in the Abell cluster A370 with redshift $z_d = 0.37$, yielding $z_s = 0.724$ (Soucail et al. 1988), clearly verified the lensing hypothesis. Many giant arcs have been discovered since, and systematic surveys have been carried out (for a recent review on giant arcs and cluster lensing, see Fort & Mellier 1994). For example, Luppino et al. (1995) found giant arcs in 8 out of 40 X-ray-selected clusters with redshift $\geq 0.15$, and the fraction of arc clusters increases with increasing X-ray luminosity. In Sect. 4.1 below I will discuss some selected results from the analysis of arcs in clusters. If a few background galaxies are so strongly distorted as to form these giant luminous arcs, it appears evident that many more background galaxies are more weakly distorted; Fort et al. (1988) were the first to discover so-called arclets in A370: images near the cluster center, still with a large axis ratio, and aligned in the tangential direction relative to the center of the cluster. Spectroscopy verified the lensing origin of the brightest of these arclets, situated at $z_s = 1.305$ (Mellier et al. 1991). Later, Tyson, Valdes & Wenk (1990) found several tens of aligned images of (presumably background) galaxies in the clusters A1689 and CL 1409+52. These discoveries then opened up the possibility to study the mass distribution in clusters, using giant arcs for the innermost part of the clusters, and the weakly distorted images in the outer parts. The finding of Kaiser & Squires (1993) of a parameter-free reconstruction of the surface mass density from observed image distortions has marked the beginning of a new and extremely promising field of research, of which some aspects and results are discussed in Sect. 4.2. The rest of this section is then devoted to other aspects of weak gravitational lensing, including the discovery of groups of galaxies through weak image distortions and magnification bias, the investigation of statistical properties of the mass distribution of galaxies, and the possibility to measure the power spectrum of density fluctuations in the universe from weak lensing.

---

[4] though arcs have been observed previously by several researchers

### 4.1 Results from giant luminous arcs

Giant arcs are the result of very strong distortions of light bundles from background sources. Such strong distortions require that the locally linearized lens mapping, described by the matrix $A$ (10), is nearly singular. In other words, giant arcs are formed near a critical curve of the cluster lens. Assuming for a moment that the cluster mass distribution is axially-symmetric, then the mass estimate as given by eq. (12) is valid, where now $\theta$ is the distance of the arc from the cluster center, and $\Sigma_{\mathrm{cr}}$ can be determined if the redshift of the arc is measured, or estimated from the color of the arc. This mass estimate is the most basic parameter one can infer from the observation of a single arc, and in the absence of additional information and assumptions, it is the only quantity that can be derived. Depending on the geometry of the cluster, this mass estimate is fairly robust; it loses its accuracy if the cluster is highly eccentric or has significant substructure. From a sample of numerically generated clusters, Bartelmann (1995a) has shown that this simple mass estimate typically overestimates the mass of the cluster within the arc distance by about 30%, however with a large scatter.

The discovery of arcs was a surprise, because it has been thought that clusters are not compact enough to produce critical curves. To understand this, consider a cluster mass profile; keeping the outer profile fixed, by reducing the core size (i.e., the length scale within which the cluster mass profile is roughly flat) the central surface mass density is increased. Clusters become critical (i.e., possess critical curves) only if the dimensionless surface mass density $\kappa$ is of order unity at the center; this requires the core size to be sufficiently small. The core radius of clusters as estimated from X-ray observations of the intracluster gas was thought to be considerably larger than needed for critical clusters. The occurrence of arcs in clusters immediately demonstrated that the core size of clusters must be small, much smaller than estimated before.

The preceding discussion has been rather vague, since the concept of a core size of a cluster is not very well defined. Basically, it is a parameter in a parametrized profile, either of the mass or the X-ray emissivity, and different parametrizations can yield different values for the core radius. However, the differences between the core size as estimated from X-ray studies (typically in excess of $100h^{-1}$ kpc) and that estimated from lensing are larger than can be easily explained as being due to semantic problems. To wit, if the mass profile of a cluster is described by an isothermal sphere with a finite core radius, in order for the cluster to be critical, the core radius must be smaller than half the Einstein radius of the cluster. Since the arc roughly traces the Einstein radius, the core radius must be smaller than half the separation of the arc from the cluster center. Given that most arcs have a separation of $\sim 20''$ from the cluster center, this argument implies core radii $\lesssim 30h^{-1}$ kpc, in marked conflict with the results from X-ray imaging. These qualitative remarks have been substantiated in detail by Miralda-Escudé & Babul (1995) who have investigated three arc clusters in detail for which X-ray observations are available. They also outlined several possible origins for the discrepancy, e.g., projection effects (which they consider unlikely), non-thermal pressure support of the intracluster gas, or a multiphase medium. When judging the seriousness of this discrepancy, one should always bear in mind the large number of assumptions entering the X-ray investigations, e.g., hydrostatic equilibrium, symmetry, isothermal gas distribution, whereas the lensing investigation is simple and purely geometrical. Recently, Waxman & Miralda-Escudé (1995) and Navarro, Frenk & White (1995) showed that the discrepancy may be reduced if the dark matter halo profile in clusters follows a universal density law, which allows an isothermal X-ray gas in hydrostatic equilibrium to develop a flat core well outside the radius where giant arcs form.

For some clusters, the observations of arcs permit a much more detailed study of their (projected) mass density. This is the case if multiple images can be identified, or if several arcs show up, or if the brightness profile of the arc permits the identification of multiply imaged components. In the cluster Cl 2137−23 ($z_\mathrm{d} = 0.313$), two arcs have been discovered (Fort et al. 1992): a tangential arc $15''\!.5$ away from the central cD galaxy and $12''$ long, and a radial arc about $5''$ long and also $5''$ away from the center of the cD galaxy. The importance of this radial arc cannot be overstated, since its position clearly indicates the turnover of the mass profile; in other words, its position directly yields the core radius of this cluster, quite independent of any details of the lens model; the resulting value is $r_\mathrm{core} = 25h^{-1}$ kpc. A detailed model of this arc system was performed by Mellier, Fort & Kneib (1993). Amazingly, an elliptical isothermal mass profile (with finite core) with the same ellipticity and orientation as the cD galaxy yields an acceptable model for the tangential and the radial arc. This model then *predicts* the locations of two additional images corresponding to the source of the tangential arc, and one additional image of the source of the radial arc, and these predicted locations are impressively close to observed arclets in the cluster (within $0''\!.6$). Hence, in this case the lens model has predictive power, and can be safely assumed to yield a realistic description of the mass distribution within the inner $\sim 15''$ of the cluster. In the cluster A370, the detailed structure of the giant arc and several multiple image candidates were used to construct a detailed mass model for this cluster (Kneib et al. 1993); also in this case, a mass model which follows closely the distribution of light yields a satisfactory fit to the observations. The giant arc in the cluster Cl 0024+16 is split up into three segments; this is caused by a clump of cluster galaxies near the arc which locally perturb the lens potential significantly. Satisfactory models of this arc system were derived by Kassiola, Kovner & Fort (1992), and a lens inversion, using techniques similar to those used for inverting radio ring images (see Sect. 3.1), has been performed by Wallington, Kochanek & Koo (1995). In this case, the mass of the perturbing galaxies can be estimated fairly accurately. As a final example, refurbished-HST images of the cluster A2218 (Kneib et al. 1995) have revealed a most amazing collection of arcs in the central parts of this cluster; together with several redshifts measured for these arcs, the most detailed mass model for the central part of a cluster currently available has been constructed.

Several heroic attempts have been made to predict the frequency of occurrence of giant luminous arcs from the observed number density of clusters, using analytical models (e.g., Wu & Hammer 1993, Bergmann & Petrosian 1993, Miralda-Escudé 1993, Grossman & Saha 1994). The results of these studies, in particular those which consider mainly spherically symmetric mass profiles for the clusters, are to be interpreted with great care, as shown by the numerical investigation by Bartelmann & Weiss (1994), and Bartelmann, Steinmetz & Weiss (1995); the probability for forming arcs in these numerically generated cluster mass profiles is substantially higher than that of more symmetric mass profiles, say with the same mass. The reason for that is that asymmetries and substructure increases the total length of the caustic curve. Another way to view this fact is that the shear is increased by substructure, such that critical curves can occur in regions where $\kappa$ is considerably less than unity (Bartelmann 1995a).

### 4.2 Cluster mass reconstruction from weak lensing

The fact that the sky is densely covered by faint galaxy images allows the statistical study of distortions of light bundles from these high-redshift sources. The basic idea here is that the shape of a galaxy image is affected by the tidal gravitational field along its corresponding light bundle. This tidal field causes a circular galaxy to form an elliptical image. Since galaxies are not round intrinsically, this effect can not be detected in individual galaxy images (except when the distortion is so strong as to lead to the formation of arcs), but since the intrinsic orientation of galaxies can be assumed to be random, a coherent alignment of images can be detected from an ensemble of galaxies. In this and the next two subsections, we shall discuss several aspects of this general idea.

If one considers the line-of-sight towards a cluster of galaxies, one can assume that the main contribution to the tidal gravitational field along light bundles corresponding to galaxies behind the cluster comes from the cluster itself, unless there are other clusters near this line-of-sight. The tidal field, or the shear, is then related to the gravitational potential $\psi$ of the cluster, as given in (11). Combining eqs.(11) and (6), and defining the complex shear $\gamma$ by $\gamma = \gamma_1 + i\gamma_2$, one finds the relation between shear and surface mass density $\kappa$ to be

$$\gamma(\boldsymbol{\theta}) = \frac{1}{\pi} \int_{\mathbb{R}^2} d^2\theta' \, \mathcal{D}(\boldsymbol{\theta} - \boldsymbol{\theta}')\, \kappa(\boldsymbol{\theta}') \quad , \tag{14a}$$

with the complex function

$$\mathcal{D}(\boldsymbol{\theta}) = \frac{\theta_2^2 - \theta_1^2 - 2i\theta_1\theta_2}{|\boldsymbol{\theta}|^4} \quad . \tag{14b}$$

Since the relation (14a) between shear and surface mass density is a convolution-type integral, it can be inverted, e.g., by Fourier methods, to yield (Kaiser & Squires 1993)

$$\kappa(\boldsymbol{\theta}) = \frac{1}{\pi} \int_{\mathbb{R}^2} \mathrm{d}^2\theta' \, \mathcal{R}\mathrm{e}\big[\mathcal{D}^*(\boldsymbol{\theta} - \boldsymbol{\theta}') \, \gamma(\boldsymbol{\theta}')\big] + \kappa_0 \quad , \tag{15}$$

where the asterisk denotes complex conjugation, and $\mathcal{R}\mathrm{e}(x)$ is the real part of the complex variable $x$. Hence, if the tidal field $\gamma$ can be measured, the surface mass density of the cluster can be obtained from (15) up to an overall constant. The reason for this constant to occur is that a homogeneous mass sheet does not cause any shear.

One can think of several methods to characterize the shape of a galaxy image. A convenient method is provided by using the matrix of second brightness moments,

$$Q_{ij} = \frac{\int \mathrm{d}^2\theta \, I(\boldsymbol{\theta}) \, (\theta_i - \bar{\theta}_i) \, (\theta_j - \bar{\theta}_j)}{\int \mathrm{d}^2\theta \, I(\boldsymbol{\theta})} \quad , \tag{16}$$

where $I(\boldsymbol{\theta})$ is the surface brightness distribution, and $\bar{\boldsymbol{\theta}}$ is the center of light of the galaxy image, defined such that $\int \mathrm{d}^2\theta \, I(\boldsymbol{\theta}) \, (\boldsymbol{\theta} - \bar{\boldsymbol{\theta}}) = 0$. Defining in analogy the tensor of second brightness moments $Q_{ij}^{(\mathrm{s})}$ of the intrinsic brightness distribution of the galaxies, one finds from the lens equation (5) and the conservation of surface brightness, $I(\boldsymbol{\theta}) = I^{(\mathrm{s})}(\boldsymbol{\beta}(\boldsymbol{\theta}))$ that $Q^{(\mathrm{s})} = A \, Q \, A$, where $A$ is given by (10).

In the following, we shall for simplicity restrict our attention to non-critical clusters only, i.e., we shall assume that $\det A > 0$ everywhere. The reader is referred to Schneider & Seitz (1995) and Seitz & Schneider (1995a) for the treatment of critical clusters. One then defines the complex ellipticity of an image as

$$\epsilon = \frac{Q_{11} - Q_{22} + 2\mathrm{i}Q_{12}}{Q_{11} + Q_{22} - 2\sqrt{Q_{11}Q_{22} - Q_{12}^2}} \quad , \tag{17}$$

and correspondingly the ellipticity $\epsilon^{(\mathrm{s})}$ of the intrinsic brightness profile of the galaxy in terms of $Q_{ij}^{(\mathrm{s})}$. For example, if an image has elliptical contours of axis ratio $r \leq 1$, then $|\epsilon| = (1-r)/(1+r)$. From the relation $Q^{(\mathrm{s})} = A \, Q \, A$ one then derives the transformation between intrinsic and observed ellipticity (Schneider 1995)

$$\epsilon^{(\mathrm{s})} = \frac{\epsilon - g}{1 - g^*\epsilon} \quad , \tag{18}$$

where

$$g = \frac{\gamma}{1 - \kappa} \tag{19}$$

is the (complex) reduced shear. Finally, averaging over a set of galaxy images, together with the assumption that the intrinsic ellipticity distribution is isotropic, so that $\langle \epsilon^{(\mathrm{s})} \rangle = 0$, one finds that

$$g = \langle \epsilon \rangle \quad . \tag{20}$$

Several comments have to made at this point:

(a) The definition (16) of the quadrupole moments cannot be applied to real images, as the integration extends to infinity. In order not to be completely dominated by noise, a weighting function has to be included in the integrals. However, with an angle-dependent weight function, the relation between $Q$ and $Q^{(s)}$ no longer has a simple form and is only approximately given by $Q^{(s)} = A\,Q\,A$; the deviations from this law depend on the intrinsic brightness profile of the source and the weighting function. Even worse is the effect of seeing and an anisotropic point-spread-function (PSF), in particular if the latter is not known very precisely. Several methods to deal with these complications have been discussed in the literature (e.g., Bonnet & Mellier 1995; Kaiser, Squires & Broadhurst 1995). In particular, a calibration of the relation between $\epsilon$ and $\epsilon^{(s)}$ is obtained from numerical simulations and from applying these methods to degraded HST images. It is clear that HST images with their unprecedented angular resolution are best suited for this kind of work, and that ground-based images are much more difficult to analyse. Future ground-based observations will make use of the calibration that can be obtained from HST images, in particular if an HST field is centered on the ground-based image.

(b) The fact that the observable $g$ has to be obtained from averaging over an ensemble of galaxy images implies that this method has a finite resolution. I.e., the averaging process is performed over the galaxy images within a certain smoothing length from the point of interest. Several methods of smoothing have been discussed (Kaiser & Squires 1993, Seitz & Schneider 1995a); we prefer smoothing with Gaussian weights. Since the number of images over which the average is perfomed is finite, the relation $\langle \epsilon^{(s)} \rangle = 0$ is not strictly valid due to the finite width of the intrinsic ellipticity distribution; only the expectation value of $\epsilon^{(s)}$ vanishes. The smoothing length need not be kept constant, but can be adapted to the local 'strength of the signal'.

(c) It is clear from (20) that only the reduced shear is an observable, but not the shear itself as needed in the inversion equation (15). If the lens is weak in the sense $\kappa \ll 1$, then $g \approx \gamma$, and (15) can be applied directly. In general, one can replace $\gamma$ in (15) by $(1-\kappa)g$, which then yields an integral equation for $\kappa(\boldsymbol{\theta})$. As shown in Seitz & Schneider (1995a), this integral equation can be easily solved in a few iteration steps. If this nonlinear correction is taken into account, then $\kappa(\boldsymbol{\theta})$ is no longer determined up to an overall additive constant as implied by (15), but there exists a global invariance transformation (Schneider & Seitz 1995)

$$\kappa(\boldsymbol{\theta}) \to \lambda \kappa(\boldsymbol{\theta}) + (1-\lambda) \quad , \tag{21}$$

which leaves all image shapes invariant. Note that this invariance transformation is the same as the mass sheet degeneracy discussed in Sect. 3.2. Of course, the allowed values of $\lambda$ are restricted by the requirement that the resulting mass distribution is non-negative. Hence, this constraint always allows to obtain a lower limit on the mass. An alternative way to obtain a lower limit to the mass inside circular apertures has been discussed by Kaiser (1995a) – the so-called aperture densitometry – which also allows a rigorous estimate of the uncertainty of this lower limit. Also, if the data field is sufficiently large, one might expect that $\kappa$ decreases to near zero at the boundary of the field, which then yields a plausible range for $\lambda$; this in fact is one of the arguments to demand wide-angle fields.

(d) The integral in (15) extends over the whole sky; on the other hand, data are given only on a finite data field (CCD field) $\mathcal{U}$. If the field $\mathcal{U}$ is not sufficiently large, and the contributions of the integral (15) from outside the data field are neglected, the estimate of the surface mass density is no longer unbiased, but boundary artefacts occur. Kaiser (1995a) noticed that there exists a *local* relation between the gradient of $\kappa$ and certain combinations of first derivatives of the shear components (which is due to the fact that both of these quantities are third derivatives of the deflection potential $\psi$). Performing averages over line integrations of this local relation allows the construction of unbiased finite-field inversion formulae (Schneider 1995, Kaiser et al. 1995, Bartelmann 1995c, Seitz & Schneider 1995b). In the latter of these papers, an inversion formula has been derived which filters out a particular noise component in the data which is readily identified as such, and a quantitative comparison with other inversion formulae has been performed.

(e) The transformation (21) leaves all image shapes invariant, but affects the magnification, $\mu \to \mu/\lambda^2$. Hence, this invariance transformation can be broken if the magnification can be measured. Two possibilities have been mentioned in the literature: Broadhurst, Taylor & Peacock (1995) noticed that the magnification effect changes the local number density of galaxy images (see footnote 3), $n(S) = n_0(S/\mu)/\mu$, where $n(S)$ are the cumulative number counts, and $n_0(S)$ are the counts in the absence of lensing. Assuming a local power law, $n_0(S) \propto S^{-\alpha}$, then $n(S)/n_0(S) = \mu^{\alpha-1}$. The blue galaxy counts have $\alpha \approx 1$, and so no magnification bias effect is observable. However, counts in the red have a flatter slope, $\alpha \approx 0.75$, and a number density decrease should be seen in regions of high magnifications. The number counts of galaxies with a red color has an even flatter slope, and the magnification effects become stronger. Indeed, this effect has been clearly seen in the cluster A1689 (Broadhurst 1995). The magnification effect also changes the redshift distribution at fixed apparent magnitude. Bartelmann & Narayan (1995) noticed that individual galaxy images become apparently brighter, at fixed surface brightness. Assuming a sufficiently tight intrinsic magnitude - surface brightness relation, the magnification can be obtained locally. The additional information coming from the magnification effects cannot be incorporated easily in a direct inversion formula such as (15), and there are two possibilities to make use of it: one could obtain the surface mass distribution from a direct inversion, such as (15), and use the magnification information afterwards to fix the transformation parameter $\lambda$ in (21). Or, one could use a reconstruction method which takes into account the *local* magnification information. One possibility for the latter is a maximum-likelihood approach (Bartelmann et al. 1995) for the reconstruction of the deflection potential $\psi$.

(f) We have implicitly assumed that all sources have the same redshift, i.e., that the critical surface mass density $\Sigma_{\mathrm{cr}}$ is the same for all sources. This assump-tion is not too bad if the cluster is at a sufficiently low redshift, since then the ratio $D_{\rm ds}/D_{\rm s}$ can be assumed constant for faint galaxies. In general, however, the redshift distribution of galaxies has to be taken into account. In the weak lensing regime ($\kappa \ll 1$, $|\gamma| \ll 1$), only the mean value of $D_{\rm ds}/D_{\rm s}$ enters the reconstruction. The non-linear case is more complicated (Seitz & Schneider 1996) and requires the functional form of the redshift distribution. On the other hand, this dependence may also allow to obtain constraints on the redshift distribution of the faintest galaxies. Alternatively, Bartelmann & Narayan (1995) pointed out that the expected strong dependence of surface brightness on the redshift of galaxies, together with the dependence of the lensing strength on source redshift, may allow to determine the redshift distribution of galaxies by studying the variation of lensing strength (i.e., mean ellipticity) as a function of surface brightness. Also, the comparison of lens reconstruction of clusters at different redshifts might allow conclusions about the redshift distribution as a function of magnitude (Smail, Ellis & Fitchett 1994).

**Fig. 2.** The WFPC2 image of the cluster Cl0939+4713 (A851); North is at the bottom, East to the right. The coordinates are in arcseconds. The cluster center is located at about the upper left corner of the left CCD, a secondary maximum of the bright (cluster) galaxies is seen close to the interface of the two lower CCDs, and a minimum in the cluster light is at the interface between the two right CCDs. In the lensing analysis, the data from the small CCD (the Planetary Camera) were not used

The cluster construction method described above has been applied to several clusters. Fahlman et al. (1994) analyzed the shear field of the cluster MS1224 and obtained a mass-to-light ratio of $\sim 800h$, where $h$ is the Hubble constant in units of 100 km/s/Mpc; in particular, the mass derived is much larger than that obtained from a virial analysis. For the cluster A1689, an $M/L$-ratio of about $450h$ was found by two independent groups (Kaiser 1995b; Tyson & Fischer 1995). A similar value for the $M/L$-ratio was found for two clusters by Smail et al. (1995).

We (Seitz et al. 1995) have recently analyzed the 'weak' lensing effects in the cluster Cl0939+4713 (A851), using WFPC2 data (Dressler et al. 1994). Since the WFPC2 field is fairly small, we have data only in the center of the cluster, where the lensing is not weak. Also, the small field requires the use of an unbiased finite-field inversion technique, and we used the one derived in Seitz & Schneider (1995b). Fig. 2 shows the WFPC2 image of the cluster, and the reconstructed mass distribution, together with results from a bootstrapping analysis, is shown in Fig. 3. From the latter figure, one infers that the reconstruction yields basically four significant features in the mass map: a maximum close to the position where the cluster center is predicted from optical observations, a secondary maximum roughly in the lower right CCD, an overall gradient in the lower two CCDs increasing 'to the left', and a pronounced minimum at the interface between the two right CCDs. Comparing these features with the image (Fig. 2) one sees that the maximum is clearly visible in the bright (cluster) galaxies, but also the secondary maximum and the minimum in the light distribution. In addition, the two maxima may be traced by the X-ray emission, as indicated by the ROSAT PSPC-map. Hence, in this cluster we have strong evidence of significant substructure in the mass, and that the light distribution on average follows this substructure. It will be interesting to compare the mass map with a detailed HRI map which will be obtained soon (S. Schindler, private communication). The $M/L$-ratio of the cluster within the WFC field depends on the assumed redshift distribution of the background galaxies. Assuming that the mean redshift of galaxies with $24 \leq R \leq 25.5$ is about unity, we find that $M/L \sim 200h$, a value significantly lower than for, e.g., MS1224. However, this is not too surprising, since A851 is the highest-redshift cluster in the Abell catalog which clearly biases towards high optical luminosity. In this cluster, we also have detected the magnification effect discussed above, which has allowed us to obtain not only a strict lower limit on the mass inside the data field, but also to obtain an estimate of the mass, which led to the above value for the $M/L$-ratio. Note, however, that this mass calibration is uncertain due to the fact that an (unknown) fraction of the faint galaxies are cluster members which renders the estimate of the magnification effect uncertain.

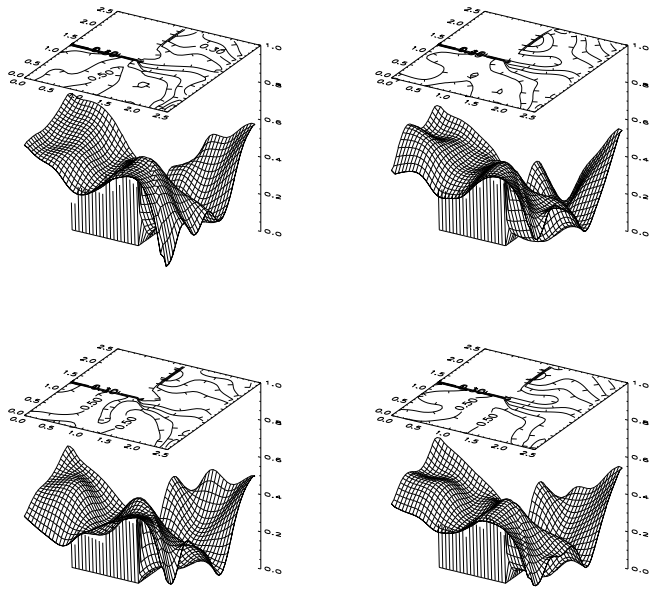

**Fig. 3.** The lower right panel shows the reconstructed mass distribution of A851, assuming a mean redshift of the $N = 295$ galaxies with $24 \le R \le 25.5$ of $\langle z \rangle = 1$. The other three panels show reconstructions obtained from the same data set via bootstrapping, i.e., selecting randomly (with replacement) $N = 295$ galaxies from the galaxy sample. The similarity of these mass distributions shows the robust features of the reconstruction, i.e., a maximum, a secondary maximum, an overall gradient, and a pronounced minimum; these features can be compared with the light distribution as shown in Fig. 2

### 4.3 Magnification effects in high-redshift QSO samples

The magnification bias which has been discussed in footnote 3, can affect the number counts of objects, provided the optical depth (or lensing probability) is sufficiently large, and the number counts of these sources are sufficiently steep. Whereas there has been a long debate of estimating the importance of this magnification bias on QSO counts, it now appears that the counts are not dramatically changed by lensing (for references and a detailed discussion, see Sect. 12.5 of SEF). Nevertheless, the fraction of magnified sources in a flux-limited sample can still be appreciable. A sign of a magnification bias could be found if high-redshift QSOs were associated with potential lenses along their lines-of-sight. Such associations have been found: on scales of a few arcseconds, several claims have been made of a detection of a statistically significant overdensity of galaxies around high-redshift QSOs (for references, see Sect. 12.3 of SEF), though the situation is not without controvercies (Wu 1996, and references therein).

Here I want to concentrate on associations on much larger scales: Tyson (1986) and Fugmann (1988, 1989) discovered a statistically significant overdensity of galaxies around high-redshift quasars on an angular scale of about one arcminute (but see Fried 1992 for a negative result). Later, Fugmann (1990) started a series of investigations to search for an overdensity of foreground matter near the lines-of-sight to high-redshift radio quasars from the 1-Jy catalog on scales of ten arcminutes and larger. Indeed, a statistically significant overdensity of galaxies from the Lick catalog (Fugmann 1990, Bartelmann & Schneider 1993), the IRAS catalog (Bartelmann & Schneider 1994, Bartsch, Schneider & Bartelmann 1996), and the APM catalog (Benitez & Martinez-Gonzalez 1995), with clusters from the Zwicky (Seitz & Schneider 1995c) and Abell (Wu & Han 1995) catalogs, and with diffuse X-ray emission from the ROSAT All Sky Survey (Bartelmann, Schneider & Hasinger 1995) were found. Further evidence for large-scale associations has come from other QSO samples (Rodrigues-Williams & Hogan 1994; Hutchings 1995). If these associations were to be explained by a lensing effect, then the lenses cannot be individual galaxies, whose 'lens-scale' is only at most a few arcseconds, but groups, clusters, or even larger-scale structures must be responsible; in fact, such an explanation works at least qualitatively (Bartelmann 1995b). If there are indeed large-scale matter overdensities in the lines-of-sight to these QSOs, they might cause a systematic distortion of background galaxies. This was the motivation for Fort et al. (1995) to image the faint galaxies around several high-redshift 1-Jy QSOs. For several of them, they obtained clear evidence for a coherent shear pattern around these QSOs, which can also be spatially related to local concentrations of faint galaxies. These concentrations may indicate the presence of a group or a cluster, but they are so faint optically that they would not appear in any cluster catalog. What this might suggest is that there exists a population of clusters with a much larger mass-to-light ratio than those clusters which are selected because of their high optical luminosity, i.e., which appear in optically-selected cluster catalogs. If these findings are confirmed (e.g., by HST observations), one has found a way to obtain a *mass-selected sample* of clusters and/or groups.

### 4.4 Galaxy-galaxy lensing

The shear field around clusters is sufficiently strong to measure their mass distributions – see Sect. 4.2. One can easily show that, assuming an isothermal mass profile, the 'detection efficiency' of a lens scales like $\sigma^4$, where $\sigma$ is the velocity dispersion. This scaling then implies that individual galaxies are too weak for their presence to be detected in their shear field[5], but one should be able to detect this effect from a large ensemble of galaxies, if the signals from the individual galaxies are added statistically. The signal one would expect is a slight tangential alignment of background galaxies relative to the direction connecting this background galaxy with a near foreground galaxy.

---

[5] assuming a number density of 50 galaxies/arcmin$^2$, the minimum velocity dispersion for which a 3-$\sigma$ detection would be possible is about 350 km/s (Miralda-Escudé 1991, Schneider & Seitz 1995).

Tyson et al. (1984) have investigated this effect using $\sim 60000$ galaxies; they obtained a null result. More recently, Brainerd, Blandford & Smail (1995) have analyzed a deep field; they have divided their galaxy sample into 'foreground' and 'background' galaxies, according to the optical magnitudes, and then studied the angle between the major axis of the background galaxy and the line connecting the background galaxy with the nearest foreground galaxies. The distribution of this angle shows a deficit at small angles, and an excess at large angles, indicating the expected tangential alignment. Since an accurate measurement of image ellipticities from the ground is very difficult, only galaxies brighter than $r = 24$ were used; the effect disappears for fainter galaxies, which most likely shows the effect of the PSF on small images. Brainerd et al. have then simulated data, treating galaxies as truncated isothermal spheres, and distributing them in redshift, and they showed that the effect they observe is in accordance with expectations from their modelling. Recalling that this effect was detected (at a 3-$\sigma$ level) with 'only' 506 'background' galaxies, it appears that one can use galaxy-galaxy lensing as a tool to investigate statistically the mass distribution in galaxies, since larger samples will become available soon (also, ground-based images with a smaller and/or more stable PSF will allow the use of fainter galaxies). Schneider & Rix (1995) have proposed a maximum likelihood method for the analysis of galaxy-galaxy lensing, which is very sensitive to the characteristic velocity dispersion of the galaxies, and which can also yield significant lower bounds on the halo size of galaxies.

### 4.5 Lensing by the large-scale structure

The cosmological density fluctuations out of which the structure in the universe has formed (at least in the conventional model of gravitational instability – which has received impressive support from the detection of microwave background fluctuations by COBE) can also distort the images of high-redshift galaxies. The corresponding distortions have been calculated by Blandford et al. (1991) and Kaiser (1992 and references therein), and are expected to be small; nevertheless, depending on the cosmological model, these distortions are measureable in principle, either by averaging the ellipticity of galaxy images over large fields, or by considering the two-point correlation function of galaxy ellipticities on large scales. If such an effect can be measured, it will allow a direct measurement of the power spectrum of the density fluctuations on the appropriate scales, very much like COBE has done. What is important to note is that the power spectrum of the density fluctuations in cosmogonies is normalized either by the amplitude of fluctuations in the microwave background, or by rms variations of galaxy numbers in 'big volumes'. Both of these normalizations are such that *relative density fluctuations* $\delta\rho/\rho$ are normalized. However, the lensing effect depends of $\delta\rho$, and not on the ratio $\delta\rho/\rho$. This implies that the gravitational distortion of images of background galaxies is proportional to the mean cosmic density $\Omega$ (Villumsen 1995a).

The same data from which galaxy-galaxy lensing was detected by Brainerd et al. (1995) have been used to search for the 'cosmic shear'; keeping in mind the difficulties to measure accurate ellipticities of very faint images from the ground, it is not surprising that Mould et al. (1994) did not find a statistically significant shear signal on a field of $4\rlap.'8$ radius. Using the same data, but a different method for analyzing the image ellipticities (basically, giving less weight to 'small' images, which are most contaminated by the PSF), Villumsen (1995b) obtained a shear signal with a formal 5-$\sigma$ significance. Further observations are needed to confirm this result; as mentioned before, the observations are very difficult to carry out, and the expected effects are so small that even tiny systematical effects which escape detection can mimic a significant detection.

## 5 Outlook

Predicting the future is a dangerous business; however, it is easy to foresee that the current developments in observational astronomy will continue to increase the usefulness of gravitational lensing for studying the universe. Concerning strong lensing, new big lens surveys, such as the CLASS survey (see, e.g., Myers et al. 1995), will allow to set much stronger constraints on the density of galaxy-mass objects in the universe. Hopefully, some of the newly discovered multiply-imaged systems will turn out to be useful for determining the Hubble constant. MACHO-type searches for compact objects in our Galaxy will continue and expand, allowing to get stronger constraints on the density of compact objects in our halo, and to measure the mass distribution in the central part of the Galaxy. Concerning weak lensing, we have just scratched the surface. On the observational side, wide-field cameras and imaging with 8m-class telescopes will dramatically increase the rate and quality of data, allowing surveys for dark matter concentrations. The refurbishment of the HST has enabled images of faint galaxies with unprecedented image quality and resolution. These images, together with new theoretical developments, will allow us to understand better the relation between observed image shapes and the true image shapes, before degradation with a PSF. The combination of dark matter maps from weak lensing and X-ray and dynamical studies of clusters will yield fresh insight into the structure, dynamics, and history of these systems. If the systematic effects of ground-based imaging can be understood sufficiently well, we might be able to obtain the cosmic density and the power spectrum of density fluctuations directly from lensing.

I would like to thank M. Bartelmann for carefully reading this manuscript. This work was supported by the "Sonderforschungsbereich 375-95 für Astro–Teilchenphysik" der Deutschen Forschungsgemeinschaft.

# References

Alcock, C. et al. 1993, Nat 365, 621.
Alcock, C. et al. 1995, Phys. Rev. Lett. 174, 2867.
Aubourg, E. et al. 1993, Nat 365, 623.
Augusto, P., Wilkinson, P.N. & Browne, I.W.A. 1996, in: *Astrophysical applications of gravitational lensing*, C.S. Kochanek & J.N. Hewitt (eds.), Kluwer: Dordrecht, p. 399.
Bartelmann, M. 1995a, A&A 299, 11.
Bartelmann, M. 1995b, A&A 298, 661.
Bartelmann, M. 1995c, A&A 303, 643.
Bartelmann, M. & Narayan, R. 1995, ApJ 451, 60.
Bartelmann, M., Narayan, R., Seitz, S. & Schneider, P. 1995, ApJ (submitted).
Bartelmann, M. & Schneider, P. 1993, A&A 271, 421.
Bartelmann, M. & Schneider, P. 1994, A&A 284, 1.
Bartelmann, M., Schneider, P. & Hasinger, G. 1994, A&A 290, 399.
Bartelmann, M., Steinmetz, M. & Weiss, A. 1995, A&A 297, 1.
Bartelmann, M. & Weiss, A. 1994, A&A 287, 1.
Bartsch, A., Schneider, P. & Bartelmann, M. 1996, A&A (submitted).
Benitez, N. & Martinez-Gonzalez, E. 1995, ApJ 448, L89.
Bergmann, A.G. & Petrosian, V. 1993, ApJ 413, 18.
Bernstein, G.M., Tyson, J.A. & Kochanek, C.S. 1993, AJ 105, 816.
Blaes, O.M. & Webster, R.L. 1992, ApJ 391, L63.
Blandford, R.D. & Narayan, R. 1992, ARA&A 30, 311.
Blandford, R.D., Saust, A.B., Brainerd, T.G. & Villumsen, J.V. 1991, MNRAS 251, 600.
Bonnet, H. & Mellier, Y. 1995, A&A 303, 331..
Brainerd, T.G., Blandford, R.D. & Smail, I. 1995, ApJ, in press.
Broadhurst, T.J. 1995, in: *Dark matter*, AIP Conf. Proc. 336, eds. S.S. Holt & C.L. Bennett (New York: AIP).
Broadhurst, T.J., Taylor, A.N. & Peacock, J.A. 1995, ApJ 438, 49.
Canizares, C.R. 1982, ApJ 263, 508.
Chen, G.H., Kochanek, C.S. & Hewitt, J.N. 1995, ApJ 447, 62.
Corbett, E.A., Browne, I.W.A. & Wilkinson, P.N. 1996, in: *Astrophysical applications of gravitational lensing*, C.S. Kochanek & J.N. Hewitt (eds.), Kluwer: Dordrecht, p. 37.
Dalcanton, J.J., Canizares, C.R., Granados, A., Steidel, C.C. & Stocke, J.T. 1994, ApJ 424, 550.
Dressler, A., Oemler, A., Butcher, H. & Gunn, J.E. 1994, ApJ 430, 107.
Fahlman, G., Kaiser, N., Squires, G. & Woods, D. 1994, ApJ 437, 56.
Fort, B., Le Fèvre, O., Hammer, F. & Cailloux, M. 1992, ApJ 399, L125.
Fort, B. & Mellier, Y. 1994, A&AR 5, 239.
Fort, B., Mellier, Y., Dantel-Fort, M., Bonnet, H. & Kneib, J.-P. 1995, A&A, in press.
Fort, B., Prieur, J.L., Mathez, G., Mellier, Y. & Soucail, G. 1988, A&A 200, L17.
Fried, J.W. 1992, A&A 254, 39.
Fugmann, W. 1988, A&A 204, 73.
Fugmann, W. 1989, A&A 222, 45.
Fugmann, W. 1990, A&A 240, 11.
Gorenstein, M.V., Falco, E.E. & Shapiro, I.I. 1988, ApJ 327, 693.

Grossman, S.A. & Saha, P. 1994, ApJ 431, 74.
Haarsma, D.B., Hewitt, J.N., Burke, B.F. & Lehár, J. 1996, in: *Astrophysical applications of gravitational lensing*, C.S. Kochanek & J.N. Hewitt (eds.), Kluwer: Dordrecht, p. 43.
Hartwick, F.D.A. & Schade, D. 1990, ARA&A 28, 437.
Hawkins, M.R.S. & Véron, P. 1993, MNRAS 260, 202.
Houde, M. & Racine, R. 1994, AJ 107, 466.
Hutchings, J.B. 1995, AJ 109, 928.
Jaroszyński, M., Wambsganss, J. & Paczyński, B. 1992, ApJ 396, L65.
Kaiser, N. 1992, ApJ 388, 272.
Kaiser, N. 1995a, ApJ 439, L1.
Kaiser, N. 1995b, preprint (astro-ph/9509019).
Kaiser, N. & Squires, G. 1993, ApJ 404, 441.
Kaiser, N., Squires, G. & Broadhurst, T. 1995, ApJ 449, 460.
Kaiser, N., Squires, G., Fahlman, G., Woods, D. & Broadhurst, T. 1995, preprint.
Kassiola, A., Kovner, I. & Blandford, R.D. 1991, ApJ 381, 6.
Kassiola, A., Kovner, I. & Fort, B. 1992, ApJ 400, 41.
Kneib, J.-P., Ellis, R.S., Smail, I., Couch, W.J. & Sharples, R.M. 1995, preprint (astro-ph/9511015).
Kneib, J.-P., Mellier, Y., Fort, B. & Mathez, G. 1993, A&A 273, 367.
Kochanek, C.S. 1993a, ApJ 417, 438.
Kochanek, C.S. 1993b, ApJ 419, 12.
Kochanek, C.S. 1995a, ApJ 445, 559.
Kochanek, C.S. 1995b, preprint (astro-ph/9510077).
Kochanek, C.S., Blandford, R.D., Lawrence, C.R. & Narayan, R. 1989, MNRAS 238, 43.
Kochanek, C.S. & Narayan, R. 1992, ApJ 401, 461.
Luppino, G.A., Gioia, I.M., Hammer, F., Le Fèvre, O. & Annis, J. 1995, ApJ (in press).
Lynds, R. & Petrosian, V. 1986, BAAS 18, 1014.
Maoz, D. & Rix, H.-W. 1993, ApJ 416, 425.
Mellier, Y., Fort, B., Soucail, G., Mathez, G. & Cailloux, M. 1991, ApJ 380, 334.
Mellier, Y., Fort, B. & Kneib, J.-P. 1993, ApJ 407, 33.
Miralda-Escudé, J. 1991, ApJ 370, 1.
Miralda-Escudé, J. 1993, ApJ 403, 497.
Miralda-Escudé, J. & Babul, A. 1995, ApJ 449, 18.
Mould, J., Blandford, R., Villumsen, J., Brainerd, T., Smail, I., Small, T. & Kells, W. 1994, MNRAS 271, 31.
Myers, S.T. et al. 1995, ApJ 447, L5.
Navarro, J.F., Frenk, C.S. & White, S.D.M. 1995, ApJ (submitted).
Paczyński, B. 1986, ApJ 304, 1.
Paczyński, B. 1991, ApJ 371, L63.
Paczyński, B. 1996, ARA&A (in press).
Patnaik, A.R., Browne, I.W.A., King, L.J., Muxlow, T.W.B., Walsh, D. & Wilkinson, P.N. 1993, MNRAS 261, 435.
Patnaik, A.R., Garrett, M.A., Polatidis, A. & Bagri, D. 1996, in: *Astrophysical applications of gravitational lensing*, C.S. Kochanek & J.N. Hewitt (eds.), Kluwer: Dordrecht, p. 405.
Patnaik, A.R., Porcas, R.W. & Browne, I.W.A. 1995, MNRAS 274, L5.
Pelt, J., Hoff, W., Kayser, R., Refsdal, S. & Schramm, T. 1994, A&A 286, 775.
Press, W.H., Rybicki, G.B. & Hewitt, J.N. 1992, ApJ 385, 404.

Rauch, K.P. & Blandford, R.D. 1991, ApJ 381, L39.
Refsdal, S. 1964, MNRAS 128, 307.
Refsdal, S. & Surdej, J. 1994, Rep. Prog. Phys. 56, 117.
Rix, H.-W., Schneider, D.P. & Bahcall, J.N. 1992, AJ 104, 959.
Roberts, D.H., Lehar, J., Hewitt, J.N. & Burke, B.F. 1991, Nat 352, 43.
Rodrigues-Williams, L.L. & Hogan, C.J. 1994, AJ 107, 451.
Schild, R. & Thomson, D.J. 1995, AJ 109, 1970.
Schneider, P. 1993, A&A 279, 1.
Schneider, P. 1995, A&A 302, 639.
Schneider, P., Ehlers, J. & Falco, E.E. 1992, *Gravitational lenses*, Springer: New York (SEF).
Schneider, P. & Rix, H.-W. 1995, preprint.
Schneider, P. & Seitz, C. 1995, A&A 294, 411.
Seitz, C., Kneib, J.-P., Schneider, P. & Seitz, S. 1995, A&A (submitted).
Seitz, C. & Schneider, P. 1995a, A&A 297, 287.
Seitz, C. & Schneider, P. 1996, A&A (submitted).
Seitz, S. & Schneider, P. 1995b, A&A, in press.
Seitz, S. & Schneider, P. 1995c, A&A 302, 9.
Seitz, S., Schneider, P. & Ehlers, J. 1994, Class. Quan. Gravity 11, 2345.
Smail, I., Ellis, R.S. & Fitchett, M.J. 1994, MNRAS 270, 245.
Smail, I., Ellis, R.S., Fitchett, M.J. & Edge, A.C. 1995, MNRAS 273, 277.
Soucail, G., Fort, B., Mellier, Y. & Picat, J.P. 1987, A&A 172, L14.
Soucail, G., Mellier, Y., Fort, B., Mathez, G. & Cailloux, M. 1988, A&A 191, L19.
Tyson, J.A. 1986, AJ 92, 691.
Tyson, J.A. & Fischer, P. 1995, ApJ 446, L55.
Tyson, J.A., Valdes, F., Jarvis, J.F. & Mills Jr., A.P. 1984, ApJ 281, L59.
Tyson, J.A., Valdes, F. & Wenk, R.A. 1990, ApJ 349, L1.
Udalski, A. et al. 1993, Acta Astro. 43, 289.
Vanderriest, C., Schneider, J., Herpe, G., Chevreton, M., Moles, M. & Wlérick, G. 1989, A&A 215, 1.
Villumsen, J.V. 1995a, MNRAS, in press.
Villumsen, J.V. 1995b, MNRAS, submitted.
Wallington, S., Kochanek, C.S. & Koo, D.C. 1995, ApJ 441, 58.
Wallington, S., Kochanek, C.S. & Narayan, R. 1995, preprint (astro-ph/9511137).
Walsh, D., Carswell, R.F. & Weymann, R.J. 1979, Nat 279, 381.
Waxman, E. & Miralda-Escudé, J. 1995, ApJ (in press).
Wu, X.-P. 1996, Fund. Cosm. Phys. (in press).
Wu, X.-P. & Hammer, F. 1993, MNRAS 262, 187.
Wu, X.-P. & Han, J. 1995, MNRAS 272, 705.
Zhao, H.S., Spergel, D.N. & Rich, R.M. 1995, ApJ 440, L13.